\input{aipcheck}

\documentclass[
    ,final            
  ]
  {aipproc}

\layoutstyle{6x9}

\begin{document}

\title{Multiwavelength Studies of Rotating \\ Radio Transients}

\classification{
97.60.Gb, 97.60.Jd, 98.70.Qy
}
\keywords      {pulsars, stars: flare, neutron, X-rays: stars}

\author{Joshua Miller}{
  address={Department of Physics, West Virginia University, Morgantown, WV 26501, USA},
  email={jmille77@mix.wvu.edu},
  thanks={This work was commissioned by the AIP}
}

\iftrue
\author{Maura McLaughlin}{
  address={Department of Physics, West Virginia University, Morgantown, WV 26501, USA},
  email={maura.mclaughlin@mail.wvu.edu},
  altaddress={Also adjunct at the National Radio Astronomy Observatory, Green Bank, WV 24944, USA}
}

\author{Nanda Rea}{address={
Institut de Ciencies de l'Espai, Campus UAB, Facultat de Ciencies, Torre C5-parell,
2a planta, 08193, Barcelona, Spain
}}

\author{Evan Keane}{
  address={Jodrell Bank Centre for Astrophysics, School of Physics and Astronomy, University of Manchester, Manchester M13 9PL, UK}
}

\author{Andrew Lyne}{
  address={Jodrell Bank Centre for Astrophysics, School of Physics and Astronomy, University of Manchester, Manchester M13 9PL, UK}
}

\author{Michael Kramer}{
  address={Jodrell Bank Centre for Astrophysics, School of Physics and Astronomy, University of Manchester, Manchester M13 9PL, UK},
  altaddress={Max Planck Institut f\"{u}r Radioastronomie, Auf dem H\"{u}gel 69, 53121 Bonn, Germany}
}

\author{Richard Manchester}{
address={ CSIRO Astronomy and Space Science, ATNF, P.O. Box 76, Epping  NSW 1710, Australia}
}

\author{Kosmas Lazaridis}{
  address={Max Planck Institut f\"{u}r Radioastronomie, Auf dem H\"{u}gel 69, 53121 Bonn, Germany}
}

\begin{abstract}
We describe our studies of the radio and high-energy properties of
Rotating Radio Transients (RRATs).
We find that the radio pulse intensity distributions are log-normal, with power-law tails
evident in two cases. For the three RRATs with coverage over a wide range of frequency,
the mean spectral index is $-1.7\pm0.1$, roughly
in the range of normal pulsars. We do not observe anomalous
magnetar-like spectra for any RRATs.
Our 94-ks {\it XMM-Newton} observation of the high magnetic field RRAT J1819$-$1458
 reveals a blackbody spectrum (kT $\sim 130$~eV) with an unusual
absorption feature at $\sim$1~keV. We find no evidence for X-ray bursts or other X-ray variability. 
We performed a correlation analysis of the X-ray photons 
with radio pulses detected in concurrent observations with the Green Bank, Effelsberg, and Parkes telescopes. We find no evidence for any correlations between radio
pulse emission and X-ray photons, perhaps suggesting that sporadicity is not due to
variations in magnetospheric particle density but to changes in beaming or coherence.
\end{abstract}

\maketitle


\section{Introduction}

Rotating Radio Transients (RRATs) were discovered
through a reanalysis of archival data from the Parkes Multibeam Pulsar Survey (PMPS)  \citep{mll+06}.
Over 30 RRATs have been discovered since that time \citep{ekl09,kle+10,bb09,dcm+09,hrk+08}
Their transient nature implies 
 a large Galactic RRAT population,
most likely outnumbering the population of normal Galactic pulsars. Unless the populations are evolutionarily related,
the implied neutron star (NS) birthrate may be inconsistent with the measured Galactic supernova rate \citep{kk08}.
In Figure~1, we compare the spin-down properties of the RRATs and other NSs. In general, RRATs have larger periods and magnetic fields than normal pulsars
\citep{mlk+09}. Some may be
dying or extreme nulling pulsars \citep{zgd07} or simply normal radio pulsars for which we only see the bright tail
of an extended pulse amplitude distribution
\citep{wsrw06}. An intriguing possibility is that their sporadicity is due to modulation
from a radiation belt \citep{lm07} or  asteroid belt \citep{li06,cs08}. It is also possible that some RRATs are transient
X-ray magnetars. An unusual glitch detected from RRAT J1819$-$1458 suggests that it is transitioning from
the magnetar to pulsar region of the $P-\dot{P}$ diagram \citep{lmk+09}. Furthermore,
{\it Chandra} observations
revealed an unusual pulsar wind nebula that is impossible to power through spin-down energy alone, suggesting a
possible magnetic origin \citep{rmg+09}. Finally, the spin-down properties of two of the RRATs place them in a region of
$P-\dot{P}$ space devoid of pulsars and close to that occupied by INSs, suggesting another possible relationship. 
\begin{figure}
  \includegraphics[height=.4\textheight,angle=270]{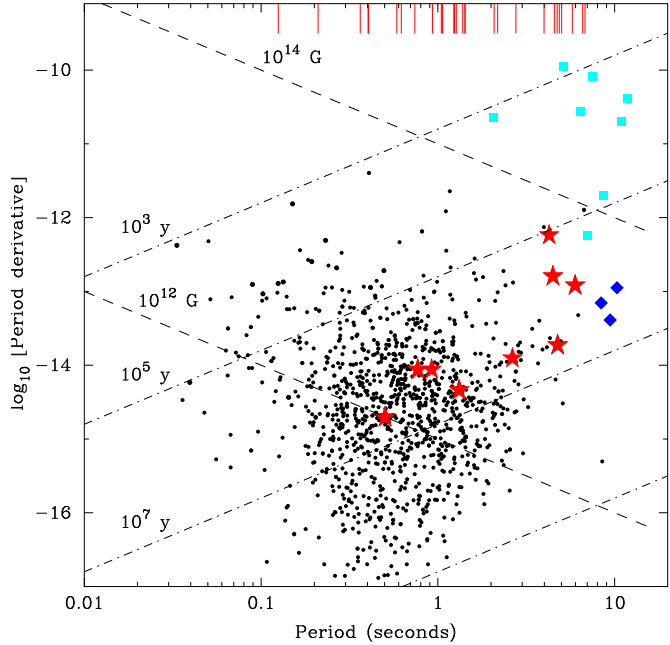}
  \caption{$P$-$\dot{P}$ diagram for RRATs (red stars),
    normal pulsars (black dots), magnetars (cyan squares), and X-ray-detected isolated neutron stars (blue diamonds).
The $\dot{P}$ values are for eight Parkes RRATs (\citep{mlk+09}, \citep{lmk+09}, \citep{bmm+11}) and one unpublished source from the 350-MHz GBT survey \citep{bmm+11}.
    Dashed lines indicate constant characteristic age and constant inferred surface dipole magnetic field strength.
Red lines at the top show the periods of the RRATs which do not yet have measured period derivatives. 
}
\end{figure}

\subsection{Pulse Amplitude Distributions and Spectra}

The pulse amplitude distributions of eight Parkes RRATs 
 are well described by log-normal probability distribution functions, with
two exceptions  where a power-law tail is also required. Log-normal distributions
are typical of pulsar single-pulse amplitude distributions \citep{joh04}, while the power-law tails are similar
to those seen for giant-pulsing pulsars and, perhaps, magnetars \citep{kbmo06,ssw+09}.
Observations at three to six radio frequencies  allow us to calculate radio spectra for the RRATs.
Fitting their pulse amplitude distributions, we find that
the fluxes follow a power-law with frequency, like normal pulsars.
For the three RRATs with observations over a wide-range ($>$ 1~GHz) of frequencies, the mean spectral index is $-1.7 \pm$0.1, consistent with those of normal pulsars. Our observations rule out flat spectra as seen for radio-emitting magnetars  \citep{crh+06,crhr07,ljk+08}.

\begin{figure}   \includegraphics[height=.3\textheight,angle=0]{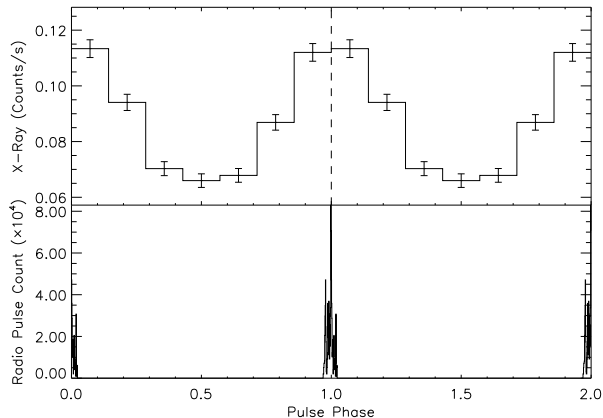}   \caption{X-ray and radio profiles of J1819$-$1458 folded using the radio ephemeris.     {\it Top:} X-ray profile consisting of seven phase bins for each rotational period     and including photons with energies 0.5~keV~$<~E~<$~2.4~keV.     The dotted line indicates the peak of the radio profile. {\it Bottom:}     Histogram of the radio pulse arrival times.
    The profile is shown twice for clarity. } \end{figure} 
\subsection{X-ray properties}

We observed the high magnetic-field RRAT J1819$-$1458 with {\it XMM-Newton} for 94~ks on 31 March 2008. These data were taken with the EPIC-PN in Full Frame mode, and the two
MOS with the central CCD in Small Window mode. Our time resolution of 73.4~ms is sufficient for studying the pulse profile. Restricting the energy range to 0.5~keV~$<~E~<$~2.4~keV,
we fit the spectrum well with an absorbed blackbody with temperature of $128\pm6$~eV and 
a cyclotron absorption line at $1.07\pm8$~eV. We can not find a good fit for the blackbody spectrum alone. Coupled with the detection of
this line in previous {\it XMM-Newton} observations \citep{mrg+07} and {\it Chandra} data \citep{rmg+09}, we are certain of its astrophysical nature. If the line is due to proton resonant cyclotron scattering, this implies a magnetic field strength of $2\times10^{14}$~G, four times as high as that measured from spin-down.
Assuming an angle between the magnetic and spin axes of 15~degrees would make the spin-down magnetic field consistent with the
cyclotron estimate. Note that while the cyclotron interpretation for the line is appealing, we are also not able to rule out an atmospheric absorption line. More careful spectral modeling and a phase-resolved analysis is required to conclusively determine
the nature of the line.

\subsection{X-ray and radio correlation analysis}

We observed J1819$-$1458 with the GBT, Parkes, and Effelsberg radio telescopes at frequencies of 2, 1.4, and 1.4~GHz, respectively, concurrently with the 94-ks {\it XMM-Newton} observation.
 We detected 6800 X-ray photons (in the energy range 0.5~keV~$<~E~<$~2.4~keV) and
931 radio pulses (165 from Parkes over $\sim$9 hours, 64 from Effelsberg over $\sim$5 hours, and 673 from the GBT over $\sim$7~hours). In Figure~2, we present the X-ray profile and the profile of the radio bursts. The two profiles align, with
the X-ray profile well-described by a sinusoid (as would be expected from thermal emission).

We searched for correlations between the radio pulses and X-ray photons by calculating the number 
arriving within some time window of each other and then repeating this analysis for simulated randomly
distributed pulses. We find no evidence for any excess X-ray emission at the times of the radio pulses
on timescales from one pulse period. 
This may imply that the RRAT radio sporadicity is due to beaming or radio coherence and not due to an 
increase in particle density in the magnetosphere, which might be expected to heat the polar cap and lead to increased
X-ray emission \citep{zgd07}.


\begin{theacknowledgments}
JM is supported by the NRAO student support program. MAM and JM are supported through
 NASA grant 
4200261589. Part of this work is based on observations with the 100-m telescope of the Max-Planck-Institut f\"ur Radioastronomie (MPIfR) at Effelsberg. The Parkes radio telescope is part of the ATNF which is funded by the Commonwealth of Australia for operation as a National Facility managed by CSIRO.
The NRAO is a facility of the NSF operated under cooperative agreement by Associated Universities, Inc.
\end{theacknowledgments}



\bibliographystyle{aipproc}   

\bibliography{journals,psrrefs,modrefs}

\IfFileExists{\jobname.bbl}{}
 {\typeout{}
  \typeout{******************************************}
  \typeout{** Please run "bibtex \jobname" to optain}
  \typeout{** the bibliography and then re-run LaTeX}
  \typeout{** twice to fix the references!}
  \typeout{******************************************}
  \typeout{}
 }

\end{document}
